\documentstyle{mn}
\input epsf
\def\plotone#1{\centering \leavevmode
\epsfxsize=\columnwidth \epsfbox{#1}}

\title[Supermassive Black Holes in Elliptical Galaxies]{Supermassive
Black Holes in Elliptical Galaxies: Switching from Very Bright
to Very Dim}

\author[Churazov,  Sazonov, Sunyaev, Forman, Jones and
B\"ohringer]{E.~Churazov$^{1,2}$, S.~Sazonov$^{1,2}$,  R.~Sunyaev$^{1,2}$, 
W.~Forman$^{3}$, C.~Jones$^{3}$, H.~B\"ohringer$^{4}$\\
$^1$ Max-Planck-Institut f\"ur Astrophysik, Karl-Schwarzschild-Strasse 1, 85741
Garching, Germany\\
$^2$ Space Research Institute (IKI), Profsoyuznaya 84/32, Moscow 117810, 
Russia\\
$^3$ Harvard-Smithsonian Center for Astrophysics, 60 Garden St.,
Cambridge, MA 02138 \\
$^4$ MPI f\"{u}r Extraterrestrische Physik, P.O.Box 1603, 85740
Garching, Germany
}

\pagerange{\pageref{firstpage}--\pageref{lastpage}}
\pubyear{2001}

\begin{document}
\maketitle

\label{firstpage}
\begin{abstract}
Relativistic outflows (mainly observed in the radio) are a
characteristic feature of both Galactic stellar mass black holes and
supermassive black holes (SMBHs). Simultaneous radio and X-ray
observations of Galactic sources have shown that the outflow is strong
at low accretion rates, but it weakens dramatically or disappears completely
at high accretion rates, manifesting structural changes in the
accretion flow. It is reasonable to assume that SMBHs follow the same
trend. For low luminosity SMBHs in nearby elliptical galaxies and
clusters, recent observations strongly suggest that the outflows play
the central role in keeping the gas hot (mechanical feedback). If the
outflow is quenched in SMBHs at high accretion rates similarly to the
behavior of galactic sources, then the straightforward consequence is
a relatively weak feedback of rapidly accreting SMBHs. We argue
that elliptical galaxies and their central engines should then evolve 
through two stages. Early on, the central SMBH rapidly grows by
accreting cooling gas at a near-Eddington rate with high radiative
efficiency but with weak feedback on the infalling gas. This stage
terminates when the black hole has grown to a sufficiently large mass
that its feedback (radiative and/or mechanical), despite the low gas
heating efficiency, is able to suppress gas cooling.  After that the system
switches to a stable state corresponding to passively evolving
ellipticals, when the accretion rate and radiative efficiency are very
low, but the gas heating efficiency is high and energy input from
the relativistic outflow keeps the gas hot.
\end{abstract}


%

\sloppypar

\section{Introduction}
It is now widely accepted that virtually every elliptical galaxy hosts
a supermassive black hole (SMBH) at its center with its mass tightly
correlated with the mass and stellar velocity dispersion of the galaxy
itself (e.g. Ferrarese \& Merritt 2000; Tremaine et al. 2002). During
the early stages of galaxy evolution, when large quantities of cold
gas are presumably present, these black holes accrete matter at high
rates and are observed as bright QSOs once they have grown to a
considerable mass. In contrast the central SMBHs of local elliptical
galaxies usually have very small bolometric luminosities. Recent X-ray
and radio observations suggest that, despite low observed
luminosities, these SMBHs in local ellipticals eject material outflows
that have a strong impact on the thermal state of the ambient hot gas,
which is characterized by a relatively short cooling time and should
cool and condense in the absence of an external source of energy.
These outflows usually manifest themselves as bubbles of radio
emitting plasma coincident with depressions in the X-ray images
(e.g. B\"ohringer et al. 1993; Fabian et al. 2000; McNamara et
al. 2000). The energetics of the outflows/bubbles and their evolution
have been extensively discussed (e.g. Owen, Eilek \& Kassim 2000;
Churazov et al. 2001, 2002; Br\"uggen \& Kaiser 2001; David et
al. 2001; Nulsen et al. 2002; Fabian et al. 2003; Ruszkowski,
Br\"uggen \& Begelman 2004; Birzan et al. 2004). At least in several
well studied objects the power of the outflow exceeds the observed
bolometric luminosity and is sufficient to offset the gas cooling
losses.

Radio emitting flows are also observed in X-ray binaries containing
stellar mass black holes and the physics there is believed to be
similar to that in SMBHs. For individual binaries the radio emission
has been observed to vary depending on the X-ray flux and on the
spectral state of the source (e.g. Fender \& Belloni 2004). In
particular it was found that an outflow ceases above a certain
accretion rate (e.g. Fender et al. 1999). There are indications that
active galactic nuclei (AGNs) may also exhibit a similar kind of radio
power behavior (Maccarone, Gallo \& Fender 2003). It thus seems
reasonable to assume that the outflow behavior is similar for galactic
binaries and SMBHs and consider the most straightforward implications
for elliptical galaxies and their central SMBHs. This is the purpose
of the present paper. Previously the role of feedback from growing
SMBHs on their host galaxies has been discussed by many authors using
different assumptions (e.g. Binney \& Tabor 1995; Silk \& Rees 1998;
Ciotti \& Ostriker 2001; Cavaliere, Lapi \& Menci 2002; Wyithe \& Loeb
2003; Granato et al. 2004; Murray, Quataert \& Thompson 2005; Sazonov
et al. 2005; Springel, Di Matteo \& Hernquist 2005, Begelman \& Nath,
2005).
 

\section{Black Hole Energy release in X-ray binaries and AGNs}
Simultaneous observations of several X-ray binaries, in particular
black hole candidates, in radio and X-ray bands have recently led to
the following broad picture (e.g. Gallo, Fender and Pooley 2003;
Fender \& Belloni 2004). In the so-called ``low/hard'' source state
X-ray and radio fluxes are well correlated with a functional form
$F_r\propto F^{0.7}_X$. In this state the accretion rate is low -
$\dot{M}\ll \dot{M}_{\rm Edd}$, the accretion flow is optically thin
(geometrically thick) and radiative efficiency
(luminosity/$\dot{M}c^2$) is much lower than canonical 10\%. Here
$\dot{M}_{\rm Edd}=\frac{4\pi G m_p M_{BH}}{\eta c \sigma_T}$, $M_{\rm
BH}$ is the black hole mass, $G$ gravitational constant, $m_p$ proton
mass, $c$ speed of light, $\sigma_T$ Thomson cross section, and
$\eta\approx $10\%. For individual sources (a notable example is
GX339-4) the $F_r/F_X$ correlation extends over at least three orders
of magnitude in X-ray luminosity. It can be extended by
another 7-9 orders of magnitude by adding a large sample of individual
AGNs (Maccarone et al 2003; Merloni, Heinz \& Di Matteo 2003). In this
state the kinetic power of radio emitting outflows can significantly
exceed their radiative power both in AGNs and binaries (e.g. Owen et
al. 2000; Gallo et al. 2003). The nucleus of M87 can be regarded as
the prototypical example of an AGN in the low accretion rate mode. In
M87, the energetics derived from the observed jets and radio lobes
suggests an outflow power on the order of $10^{44}$ erg~s$^{-1}$,
requiring an accretion rate of order $3\times 10^{-4}\dot{M}_{\rm
Edd}$ onto a $3\times 10^9 M_\odot$ black hole for an outflow
efficiency (power/$\dot{M}c^2$) of 10\%. On the other hand, the
current X-ray luminosity of the nucleus is less than $10^{41}$
erg~s$^{-1}$ and even after bolometric correction it remains 10--100
times lower than the outflow power.

\begin{figure}
\plotone{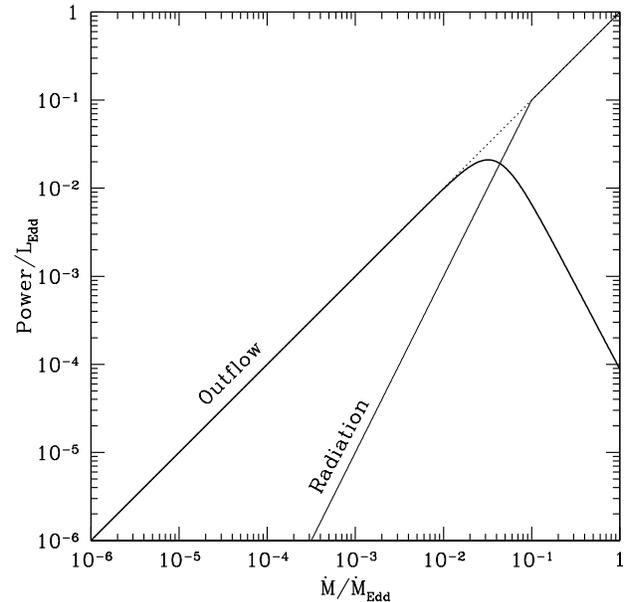}
\caption{Sketch of black hole energy release as a function of mass
accretion rate. At low accretion rates a significant fraction of
accretion power goes into an outflow (thick solid line). Above a certain
accretion rate, corresponding to the transition to the standard
accretion disk mode, the outflow power decreases. The radiative power 
(thin solid line) is instead very low at low accretion rates and 
reaches a fixed fraction ($\sim$10\%) of $\dot{M}c^2$ at accretion
rates above 0.01-0.1 of Eddington. 
\label{fig:accretion}
}
\end{figure}

For binaries the $F_r/F_X$ correlation breaks down when higher
X-ray fluxes are considered (Fender et al. 1999; Gallo et
al. 2003). This break is associated with the increased
mass accretion rate and a transition of the source to the so-called
``high/soft'' state. In this state the bulk of the observed X-ray
emission comes from a standard, optically thick and geometrically thin
accretion disk (Shakura \& Sunyaev 1973), and the spectra are
reasonably well described by the multicolor black body model. In this
state the radio emission drops by a large factor (or completely
disappears) suggesting quenching of the outflow. Once again one can
identify a similar behavior for AGNs (Maccarone, Gallo \& Fender
2003), although the trend is less clear than for binaries.

There is, therefore, good observational evidence that the energy
release via accretion onto black holes in X-ray binaries and AGNs
depends on the accretion rate qualitatively as shown in
Fig.\ref{fig:accretion}. At low rates there is a strong outflow whose
power correlates with the accretion rate. The radiative efficiency at
these rates is low, as e.g. in advection dominated flows (Esin,
McClintock \& Narayan 1997), so that the outflow dominates the output
energetics. Numerous observations of Galactic X-ray binaries have
shown that at an accretion rate of order of $\sim 10^{-2} -
10^{-1}\dot{M}_{\rm Edd}$ the sources make a spectral transition from
the low/hard to high/soft state. We use this value as the
characteristic accretion rate above which the outflow power
decreases. In the particular case of GX339-4 the radio flux has
dropped by a factor of more than 25 during the transition to the soft
state (Fender et al. 1999), which can be viewed as evidence of a
similarly strong decrease of the outflow power. In the absence of a
justifiable quantitative model of the outflow we adopt a simple
prescription (shown in Fig.\ref{fig:accretion}) according to which the
outflow power drops by large a factor when the accretion rate rises
from $\sim 10^{-2}\dot{M}_{\rm Edd}$ to $\sim \dot{M}_{\rm Edd}$.

Despite the above justification of the picture presented in
Fig.\ref{fig:accretion}, it may still be an oversimplification of the
real situation. First, as is well known, $\sim 10$\% of QSOs exhibit
strong radio emission (e.g. Ivezi\'{c} et al. 2002), indicating that
powerful outflows may sometimes be associated with rapidly accreting
SMBHs. What determines the radio loudness/quietness of QSOs remains an
open question. It is possible that apart from the instantaneous
accretion rate, other properties such as black hole spin or the
pre-history of the accretion rate may play crucial roles. An
indication that the latter may indeed be important comes from
observations in transient X-ray binaries of a short, intense radio
outburst during the transition from low/hard to high/soft state -- in
the so-called very high state -- when the accretion rate approaches
the Eddington limit (Fender, Belloni \& Gallo, 2004); such radio
flares, however, have never been observed during the reverse
transition (from high/soft to low/hard state). Importantly, it is not
yet clear whether the observed radio outbursts are associated with a
significant additional energy output or they mostly reflect a change
in the emission properties of the previously (in the low/hard state)
ejected material.
 
\section{Black Hole Feedback in Elliptical Galaxies}
We now discuss the conversion of energy released by accretion onto
SMBHs into gas thermal energy in ellipticals.

\begin{figure}
\plotone{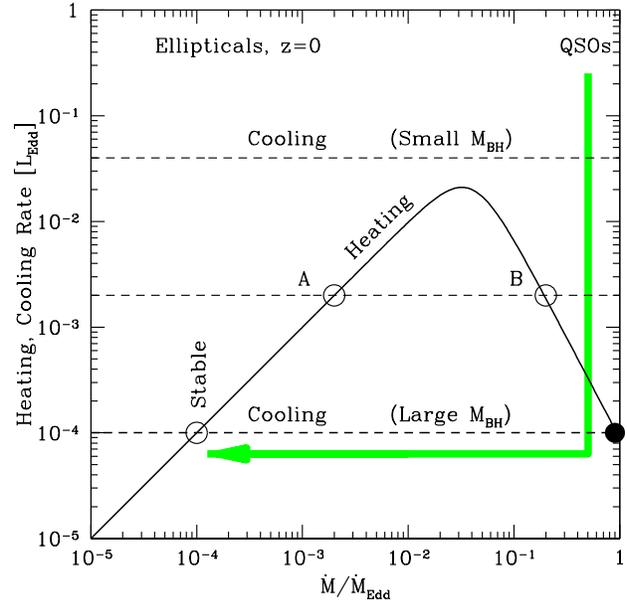}
\caption{Illustration of gas heating and cooling in elliptical
galaxies. The thick solid line shows, as a function of the SMBH
accretion rate, the heating rate due to outflow, which is
complemented/dominated by radiative heating near the Eddington
limit. Horizontal dashed lines show the gas cooling rate. The upper
cooling line represents a young galaxy in which a large amount of gas
is present and/or the black hole is small. Feedback from the black
hole is not able to compensate for gas cooling losses and the black
hole is in the QSO stage with a near-critical accretion rate, high
radiative efficiency and weak feedback. As the black hole grows it
moves down in this plot. The black solid dot marks the termination of
this stage, when the black hole is first able to offset gas cooling,
despite the low gas heating efficiency. The lower cooling line
illustrates present day ellipticals: a stable solution exists at low
accretion rates when mechanical feedback from the black hole
compensates gas cooling losses. The radiative efficiency of accretion
is very low and black hole growth is very slow.
\label{fig:feedback}
}
\end{figure}

At low accretion rates the most important form is mechanical energy input
from outflows generated by the black hole. Examples of outflow
interactions with the ISM are found in many elliptical galaxies
(e.g. Finoguenov \& Jones 2001). If the outflow is halted well within the
region occupied by the cooling gas, then the efficiency of converting the
outflow energy into the gas heating can be high, up to 100\%
(e.g. Churazov et al. 2002). For simplicity we assume
that the gas heating rate due to SMBH feedback is equal to the outflow
power. This leads to a dependence of gas heating rate on the accretion
rate (Fig.\ref{fig:feedback}) that mirrors the outflow power curve
(Fig.\ref{fig:accretion}). 

Apart from the mechanical feedback, radiative gas heating, via
photoionization and Compton scattering of radiation from the SMBH,
should be at work (Ciotti \& Ostriker 2001; Sazonov, Ostriker \&
Sunyaev 2004).  Because of low efficiency of converting luminosity
into heat the radiative feedback can be safely neglected at low
accretion rates, but it can however be important at near-Eddington
luminosities (see below). It may therefore contribute significantly
(or dominate) the high-$\dot{M}$ part of the heating curve in
Fig.\ref{fig:feedback} far from its maximum. This insures that total
feedback power (mechanical plus radiative) does not go zero near
Eddington rate even if the outflow quenches completely.

It is convenient to characterize the gas heating power $F$ through the
efficiency $\delta$ of converting accreted mass into heat: $F=\delta
c^2\dot{M}$. For subsequent analysis we assume that at very low
accretion rates (outflow regime), the efficiency is high,
$\delta_O\approx const \sim 0.1$, while close to the Eddington limit,
the efficiency is much lower, $\delta_E \ll \delta_O$.

The schematic heating curve plotted in Fig.\ref{fig:feedback} predicts
the maximum feedback power at the level of $0.02\times L_{Edd}$,
i.e. $\sim 2~10^{45} {\rm erg/s}$ for a $10^9 M_\odot$ black
hole. Comparison with the most powerful outbursts identified so far in
MS0735.6+741 (McNamara et al., 2005) and Hercules A (Nulsen et al.,
2005), both having an estimated outburst power at the level of $\sim
2~10^{46} {\rm erg/s}$, suggests that the maximum jet power can be an
order of magnitude higher than adopted in
Fig.\ref{fig:feedback}. Unless these are very exceptional systems or
their power is significantly overestimated, the efficiency of energy
extraction by jets might need to be upscaled significantly from $\sim
10\%$ adopted here and/or the heating curve peaks at a higher
accretion rate.

Consider now a gas halo feeding a black hole at its center. If the
black hole has a mass such that its maximum possible heating power
exceeds the gas losses (so that the cooling line intersects the
heating curve in Fig.\ref{fig:feedback}, e.g middle horizontal dashed
line), the system can find itself in one of two equilibrium states
(labeled {\bf A} and {\bf B}), in which cooling is balanced by
heating. We argue below that only one of these states is stable.

Consider for example that the mass accretion rate onto the black hole
is described by the Bondi formula:
\begin{eqnarray} 
 \dot{M}_{Bondi}=4\pi \lambda (GM_{\rm BH})^2 c_s^{-3}\rho \propto s^{-3/2},
\label{eqn:bondi}
\end{eqnarray}
where { $\lambda$} is a numerical coefficient that depends on the gas
adiabatic index $\gamma$ ($\lambda=0.25$ for $\gamma=5/3$),
$c_s=\sqrt{\gamma\frac{kT_e}{\mu m_p}}$ is the gas sound speed, $\rho$
is the gas mass density, and $s\propto T_e/\rho^{\gamma-1}$ is the gas
entropy index. Then it is the gas entropy near the black hole which
regulates the amount of heat supplied by the SMBH into the
gas. Consider the system near state {\bf B} (Fig.\ref{fig:feedback}),
with a heating rate slightly exceeding cooling losses. The gas 
then heats up, its entropy increases, and it expands. The gas expansion will
decrease the gas cooling losses. The Bondi accretion rate will also
decrease. But according to our feedback prescription in the vicinity
of {\bf B} (see 
Fig.\ref{fig:feedback}) the feedback power will instead increase. Thus
the imbalance between heating and cooling will increase and the system
will evolve away from state {\bf B}.

For the system in state {\bf A} (Fig.\ref{fig:feedback}), the stability
of the equilibrium depends on the specifics of energy deposition into
the gas and on the gas distribution. We can verify the stability for a
simple one-zone gas model. Consider a lump of gas with a fixed total
mass $M_{gas}$, kept at fixed pressure. The total cooling losses of
the lump $C\propto n^2 R^3 \Lambda(T) \propto n\Lambda(T)$, where $n$
is the gas density, $R$ the size of the lump, and $\Lambda(T)$ is the gas
cooling function. If we neglect changes of the cooling function with
temperature then $C \propto s^{-2/5}$. This is a less steep function
of entropy than the feedback power $\propto \dot{M}_{Bondi}\propto
s^{-3/2}$ near state {\bf A} and therefore the equilibrium will be
stable.

\subsection{A Toy Model}
We now consider the implications of our model for the evolution of elliptical
galaxies. We completely neglect stellar feedback, which can be
especially important for low mass systems. Thus the discussion below
mainly refers to massive systems with velocity dispersions
larger than $\sim$200 km/s.

We follow the picture of a quasi-steady cooling flow in an isothermal
halo (e.g. White \& Frenk 1991). We assume that the potential of the
galaxy is described by an isothermal sphere $\phi=2\sigma^2 {\rm ln}
r$ and the initial gas distribution follows the dark matter:
\begin{eqnarray}
\rho_{gas}=f_{gas}\frac{\sigma^2}{2\pi G r^2},
\label{eqn:rho}
\end{eqnarray}
where $\sigma$ is the velocity dispersion, $r$ is the distance from
the center and $f_{gas}$ is the gas mass fraction. The angular
momentum of the gas is neglected. The initial mass of the black hole
is assumed to be small enough that its feedback can be completely
neglected (situation corresponding to the upper cooling line in
Fig.\ref{fig:feedback}).  The cooling radius $r_{\rm cool}$ is set by
equality of the cooling time $t_{\rm
cool}=\frac{3/2(n_e+n_i)kT}{n_en_i \Lambda(T)}\approx
\frac{3kT_e}{n_e\Lambda(T)}$ to the age of the system $t$, where
$n_e$ and $n_i$ are the electron and ion 
densities:
\begin{eqnarray}
r_{\rm cool}=\left ( \frac{f_{gas}\Lambda(T)t}{3\pi m_{p}^2 G}\right
)^{1/2}\approx 57~  t_9^{1/2} \Lambda_{23}^{1/2}~ {\rm kpc},
\label{eqn:rcool}
\end{eqnarray}
where hereafter $\Lambda(T)= \Lambda_{23}\times 10^{-23} {\rm
erg~s^{-1}~cm^{-3}}$, $\sigma=\sigma_{200}\times 200~{\rm km/s}$,
$t=t_9\times 10^9~{\rm yr}$ and the gas fraction is set to
$f_{gas}=0.17$ (Cosmological baryon fraction).  Inside the cooling
radius the gas flows inward due to cooling losses, with the density of
the gas increasing approximately as $r^{-3/2}$ and the temperature
being approximately equal to the virial one ($T\approx 3\times
10^6\sigma_{200}^2$ K). At a radius $r_s \ll r_{cool}$ (at least for
massive systems) the cooling time becomes comparable to the sound
crossing time of the region, $t_{\rm cool}\approx r_s/c_s$:
\begin{eqnarray}
r_s=\frac{2}{9\sqrt{3}\pi^{3/2}}\frac{f_{gas}^{3/2}\Lambda(T)^{3/2}}
{\gamma m_p^3 G^{3/2} \sigma^2 t^{1/2}}\approx  1.7~
\sigma_{200}^{-2} ~ t_9^{-1/2} ~ \Lambda_{23}^{3/2} {\rm kpc}.
\label{eqn:rs}
\end{eqnarray}

In the vicinity of $r_s$ the gas cools rapidly and the temperature
drops. We assume that if a simple cooling flow solution extends down
to $r_s$ then cold gas will be present in large amounts in the system,
stars will be actively forming, and the black hole will accrete at the
Eddington rate, independently of the details. The black hole mass then
grows exponentially with the Salpeter e-folding time $t_s\approx
4\times 10^7$ yr: $M_{\rm BH}\propto e^{t/t_s}$. The mass cooling rate
of the gas is instead a decreasing function of time:
\begin{eqnarray}
\dot{M}_{\rm cool}=\frac{2}{\sqrt{3\pi}}\frac{f_{gas}^{3/2}\sigma^2
\Lambda(T)^{1/2}}{m_p G^{3/2} t^{1/2}} \approx   180~
\sigma_{200}^{2}  t_9^{-1/2} \Lambda_{23}^{1/2}~ M_\odot/{\rm yr}.
\label{eqn:mdcool}
\end{eqnarray}
White \& Frenk (1991) noted that the gas infall rate, calculated as a
time derivative of the gas mass within the virial radius
($r_{vir}=2\sigma_{km/s} (1+z)^{-3/2}$ kpc), rather than the cooling
rate may be a limiting factor. However for a redshift $z$ of 2-3 the
infall rate is not much smaller than the cooling rate (see Fig.2 of
White \& Frenk 1991). Therefore one should not expect dramatic changes
if the cooling rate is replaced by the minimum of the infall and
cooling rates, especially if massive systems are considered.  We also
note that at $z\sim$ 2-3 the cooling rate (Eq.\ref{eqn:mdcool}) is
much larger than the Eddington accretion rate (for local
$M_{BH}/\sigma$ relation).

The gas cooling losses near $r_s$ are accordingly $C\approx
3/2kT\frac{\dot{M}_{\rm cool}}{\mu mp} \propto t^{1/2}$. Additional
losses between $r_s$ and $r_{\rm cool}$ are of order of $C{\rm
ln}(r_{\rm cool}/r_s)$. With time cooling losses will decrease relative to the
(growing) heating power of the black hole accreting at the
Eddington rate. With our definition of the feedback amplitude, the black
hole accreting at the Eddington rate will be able to offset gas cooling
provided that:
\begin{eqnarray}
\delta_E c^2 \dot{M}_{\rm Edd}\approx C \approx
3/2kT\frac{\dot{M}_{\rm cool}}{\mu mp}.
\end{eqnarray}
Substituting the expression for the Eddington accretion rate ($\propto
M_{\rm BH}$) and solving for $M_{\rm BH}$ we find:
\begin{eqnarray}
M_{\rm BH}\approx \frac{\eta}{\delta_E}\frac{\sigma^4 f_{gas}^{3/2}
\Lambda(T)^{1/2} \sigma_T}{2 \sqrt{3} \pi^{3/2} c G^{5/2} m_p^2
t^{1/2}}\approx \label{eqn:mbh} \\ 
4\times 10^8  \left (
\frac{\delta_E}{10^{-5}} \right )^{-1} \sigma_{200}^4 t_9^{-1/2}
\Lambda_{23}^{1/2}~ M_\odot \nonumber
\end{eqnarray}
for an accretion radiative efficiency $\eta=0.1$. Once the black hole
has grown to this size it will be able to prevent gas cooling,
cold gas will disappear and star formation will cease. The entropy of the gas
increases dramatically and the system evolves toward a low
accretion rate solution. In this regime the accretion rate required to
produce the same heating power as before is substantially
sub-Eddington: $\dot{M}\sim \frac{\delta_O}{\delta_E}\dot{M}_{\rm Edd}$.

We can infer from equation (\ref{eqn:mbh}) that for a black hole mass
satisfying the observed $M_{\rm BH}$--$\sigma$ relation (Tremaine et
al. 2002), a balance between heating and cooling can be achieved for a
quite low feedback efficiency (power/$\dot{M}c^2$)
$\sim$$10^{-5}$. The minimum possible feedback is provided by
radiative heating, whose efficiency is ($10^{-3}$--$10^{-2}$)$\tau$
(Sazonov et al. 2005), where $\tau$ is the optical depth of the heated
gas. Using equations (\ref{eqn:rho}) and (\ref{eqn:rs}), we can
estimate the optical depth of the cooling flow external to $r_s$ as
$\tau\approx 0.004\sigma_{200}^3\Lambda_{23}^{-1}$. Furthemore, for relatively
large galaxies ($\sigma>200$~km~s$^{-1}$), the cooling flow extends
to a few kpc from the SMBH, where gas can be heated on the time scale
of SMBH growth (Sazonov et al. 2005). Therefore, radiative feedback
alone appears sufficient to suppress the cooling flow once 
the black hole accreting at a near-critical rate reaches the observed
$M_{\rm BH}$--$\sigma$ relation. Any additional channel of energy input
into the cooling gas, e.g. via relativistic jets, will only faciliate
the establishment of thermal equilibrium.

The resulting system evolution is shown in Fig.\ref{fig:feedback} as
the gray track. The black hole first accretes at the Eddington level
(vertical track) and then switches (horizontal track) to the low
accretion rate mode. The vertical track corresponds to a very high
radiation efficiency and active star formation (cold gas is 
continuously deposited), whereas in the low accretion regime the
radiative efficiency is low and little star formation takes place
in the galaxy. 

Importantly, for the black hole mass for which the thermal equilibrium
is reached, the maximum possible feedback power (maximum of the
function $F$, see Fig.\ref{fig:feedback}) significantly exceeds the
cooling losses of the gas. This has two possible implications. First,
the transition to the low accretion rate state might be followed by a
very large energy input from the SMBH and ejection of gas from the
system. Present day ellipticals are rather gas poor and thus may have
suffered significant gas loss during the transition phase. On the
other hand, the large maximum feedback power ensures that the system
will be able to find a stable equilibrium at a low accretion rate to
support a large atmosphere of hot gas even without gas ejection. The
latter case is probably relevant to cD galaxies in rich
clusters. Since the gas heating efficiency in the low accretion rate
state is high, the total mass accumulated by the SMBH over $\sim 10$
Gyr of passive evolution will not be significant: $M_{\rm BH}\approx 2\times
10^7 L_{43} \left (0.1/\delta_O
\right ) M_\odot$, where $L_{43}$ is the gas cooling losses in units
of $10^{43}$ erg~s$^{-1}$. Only for rich clusters can the black hole 
significantly increase its mass in this mode (e.g. Fabian et al. 2002).  

In rich clusters the cooling losses might exceed the maximum of the
feedback curve. The central source can then enter a QSO mode
again. The new episode of rapid accretion and black hole growth will
continue until the SMBH grows enough to switch back to the low
accretion rate mode. Such bursts of rapid accretion would be
reminiscent of the models considered by Ciotti \& Ostriker
(2001). Similar episodes of QSO-type activity could be initiated by a
merger with gas rich galaxies (as in simulations of Springel, Di
Matteo, \& Hernquist 2005) when large quantaties of cold/cool gas are
suddenly dumped onto the existing black hole.

The above model assumes that when the gas is able to cool the SMBH is
accreting at the Eddington rate. The results would be different if the
accretion rate is set by the Bondi rate from the hot phase of a
multiphase medium (along the line of reasoning suggested by Nulsen and
Fabian 2000). The thermal instability within $r_s$, i.e. where the
cooling time becomes shorter than the sound crossing time (e.g. Fall
\& Rees 1985) limits the density of the hot phase to maintain
approximate equality of these time scales. As a result the Bondi
accretion rate is limited to $\dot{M}\approx \phi
\frac{kT}{\Lambda(T)} \eta c\sigma_T
\dot{M}_{\rm Edd}\approx \phi ~ 0.3 \left(T/10^7 \right )
\Lambda_{23}^{-1} \dot{M}_{\rm Edd}$, where the dimensionless
coefficient $\phi$ of order unity depends on the detailed
assumptions (see Nulsen \& Fabian 2000). For the most massive
systems, this maximum accretion rate is not far from the Eddington
rate, while for low masses, the maximum rate may be several orders of
magnitude below Eddington. 

\section{Summary}
Guided by the analogy of SMBHs and the stellar mass black holes in
X-ray binaries, we postulated a non-monotonic dependence of the SMBH
feedback power on the mass accretion rate associated with a transition
from geometrically thick accretion flow to a geometrically thin
disk. The gas heating rate is maximal at substantially sub-Eddington
accretion rates and drops strongly for near-Eddington rates. Such a
scenario naturally leads to two stages in the evolution of a massive
elliptical galaxy: i) QSO-like nucleus and active star formation at
early epochs and ii) weak nucleus and passive evolution at the present
time.

The main difference of the picture presented here with respect to
previous work is that the phase of active black hole growth and star
formation terminates when the SMBH produces enough power
to substantially alter the thermal state of the infalling cooling gas,
rather than to immediately expell the bulk of the gas from the galaxy, which
requires much more energy (see e.g Silk \& Rees 1998; Wyithe \& Loeb
2003). In our picture, heating the gas increases its entropy and also
enhances the SMBH feedback power. The gas can (and probably will) be
expelled but this may happen over a long period of time after the end
of the QSO stage. We showed that a $M_{\rm BH}\propto \sigma^4$ dependence
with the observed normalization naturally arises for the simplest
scenario of an isothermal gas halo if the SMBH feedback efficiency in the
QSO phase is low ($\sim 10^{-5}$) and not varying significantly with
black hole mass. Furthermore, the postulated non-monotonic feedback
curve results in ``over-grown'' black holes, i.e. SMBHs capable
of supporting much more massive atmospheres of hot gas than those
present at the end of the QSO stage and afterwards.

\label{lastpage}
\end{document}